\documentclass[12pt]{article}
\usepackage{epsfig,epsf}
\begin{document}

\begin{flushright}
CERN-PH-TH/2008-094\\
SHEP-08-18\\
\end{flushright}

\begin{center}
{\Large{\bf Perturbative Estimates of the Eigenfunctions of the Non-Forward BFKL Kernel}} \\
[0.5cm]

{\large  D.A. Ross} \\ [0.3cm]

{\it Theory Division, Physics Department, CERN, CH-1211 Geneva 23, Switzerland}
\\[0.1cm]
and \\
{\it  School of Physics and Astronomy, University of Southampton}\\
{\it Highfield, Southampton SO17 1BJ, UK} \end{center}
\vspace*{3 cm}

\begin{center}
{\bf Abstract} \end{center}
We discuss, within the context of first order perturbation theory, the correction
to the NLO BFKL wavefunction for scattering processes with non-zero momentum transfer,
arising from the fact that in NLO the kernel is not covariant under conformal transformations.

\vspace*{3 cm}

\begin{flushleft}
May 2008\\
\end{flushleft}

\newpage

Recently the BFKL kernel at NLO has been presented in the dipole form \cite{ffp}. To this order,
we may write the kernel  as
\begin{equation}
{\cal K}(\mathbf{r_1},\mathbf{r_2};\mathbf{r_3}\mathbf{r_4}) \ = \
\bar{\alpha}_s{\cal K}_0(\mathbf{r_1},\mathbf{r_2};\mathbf{r_3},\mathbf{r_4},) +
\bar{\alpha}_s^2{\cal K}_1(\mathbf{r_1},\mathbf{r_2};\mathbf{r_3},\mathbf{r_4}),
\ \ \ \ \left(\bar{\alpha}_s=\frac{3\alpha_s}{\pi}\right)\label{eq1}
\end{equation}
where the kernel interpolates between a dipole with transverse coordinates $\mathbf{r_1},\mathbf{r_2}$
and a dipole with transverse coordinates $\mathbf{r_3},\mathbf{r_4}$. The rapidity dependence of
a colour singlet amplitude to NLO in rapidity, $Y$ is therefore given by the BFKL equation \cite{bfkl1,fl}
\begin{equation}
\frac{\partial}{\partial Y} f(\mathbf{r_1},\mathbf{r_2}, Y) \ = \
 \int d^2 \mathbf{r_3}d^2 \mathbf{r_4}
{\cal K}(\mathbf{r_1},\mathbf{r_2};\mathbf{r_3},\mathbf{r_4})  f(\mathbf{r_3},\mathbf{r_4}, Y).
\label{bfkl}\end{equation}
It was pointed out in \cite{lipatov86} that the leading order kernel ${\cal K}_0(\{\mathbf{r_i}\})$
possesses the following properties
\begin{enumerate}
\item $${\cal K}_0(\{\mathbf{r_i}-\mathbf{r_0}\})={\cal K}_0(\{\mathbf{r_i}\})$$
\item $${\cal K}_0(\{\Lambda \mathbf{r_i}\})= \Lambda^{-4}{\cal K}_0(\{\mathbf{r_i}\})$$
\item $${\cal K}_0(\{ 1/\mathbf{r_i}\})=  |\mathbf{r_3}|^4 \,|\mathbf{r_4}|^4 \,
 {\cal K}_0(\{\mathbf{r_i}\}).$$
\end{enumerate}
The covariance of the kernel under two-dimensional conformal transformations permits one to
write the eigenfunctions of the kernel ${\cal K}_0$,
with eigenvalues $\chi_0(\nu)$, in terms of representations of the
conformal group labeled by a real index $\nu$ \footnote{In this paper we restrict our discussion
to the case of conformal spin $n=0$.} and a transverse ``centre-of-mass'', $\mathbf{r_0}$, which we write
as
\begin{equation}
 \phi^\nu \left(\mathbf{r_1},\mathbf{r_2};\mathbf{r_0}\right)\ \equiv \ \sqrt{2} \frac{\nu}{\pi^2}
 \left[\frac{\left|\mathbf{r_{12}}\right|}
{\left|\mathbf{r_{10}}\right|\left|\mathbf{r_{20}}\right|}
 \right]^{1+2i\nu}, \label{eq3}\end{equation}
where we use the notation $\mathbf{r_{ij}}=\mathbf{r_i}-\mathbf{r_j}$.
These functions  are normalized as
\begin{eqnarray}
\langle \nu^\prime, \mathbf{r_0}^\prime | \nu, \mathbf{r_0} \rangle & \equiv &
 \int \frac{d^2 \mathbf{r_1} d^2 \mathbf{r_2}}{|\mathbf{r_{12}}|^4}
\phi^{\nu^\prime *}\left(\mathbf{r_1},\mathbf{r_2};\mathbf{r_0}^\prime \right)
\phi^{\nu}\left( \mathbf{r_1},\mathbf{r_2};\mathbf{r_0} \right)
\nonumber \\
 & = &
\delta\left(\nu-\nu^\prime\right) \delta^2\left(\mathbf{r_{00^\prime}} \right)
+\frac{2\nu}{\pi} \frac{\eta(\nu)}{\left| \mathbf{r_{00^\prime}} \right|^{2+4i\nu}}\delta\left(\nu+\nu^\prime\right),
\label{eq4} \end{eqnarray}
where $\eta$ is a pure phase given by
$$\eta(\nu) \ = \ -i 2^{4i\nu} \frac{\Gamma\left( i\nu \right) \Gamma(\left(1/2-i\nu\right)}
{\Gamma\left( -i\nu \right) \Gamma(\left(1/2+i\nu\right)}. $$ These eigenfunctions have been exploited in ref.\cite{np}
to determine the exact solution to the non-forward BFKL equation at leading order.

Provided the terms in the NLO kernel, ${\cal K}_1(\{\mathbf{r_i}\})$,
which depend on the renormalization scale, $\mu$, are absorbed into the running of the coupling
$\bar{\alpha}_s$ in front of the leading term, then the NLO kernel also obeys the
first two properties listed above. However,
as was pointed out in \cite{ffp}, the NLO kernel  does {\it not} obey the third property
and is therefore not fully conformal invariant so that the eigenfunctions of the kernel cannot be
set equal to the functions defined in (\ref{eq3}) and are consequently unknown
\footnote{It has been suggested in ref.\cite{ffp} that
it may be possible to absorb the terms which are not fully conformally invariant into 
redefinitions of the impact factors, but so far this has not been achieved.}
.

It is, however, sufficient to ascertain the solutions to the NLO
BFKL equation (\ref{bfkl}) up to order $\bar{\alpha}_s$ (relative
to the leading order) and this can be achieved by treating ${\cal
K}_1$ as a perturbation and using first-order ``rapidity
dependent'' perturbation theory. Thus we impose an initial
condition at $Y=0$
\begin{equation}
 f^\nu(\mathbf{r_1},\mathbf{r_2};\mathbf{r_0}, 0) \  = \
\phi^{\nu}\left( \mathbf{r_1},\mathbf{r_2};\mathbf{r_0} \right),  \end{equation}
and using the fact that the functions $\phi^{\nu}\left( \mathbf{r_1},\mathbf{r_2};\mathbf{r_0}\right)$
form a complete set of functions on the space $\mathbf{r_1},\mathbf{r_2}$  we may expand the
solution for non-zero rapidity  as
\begin{equation}
f^\nu(\mathbf{r_1},\mathbf{r_2};\mathbf{r_0}, Y) \
 = \ a^D(\nu,Y) \phi^{\nu}\left( \mathbf{r_1},\mathbf{r_2};\mathbf{r_0} \right)
+ \bar{\alpha}_s \int d\nu^\prime d^2\mathbf{r_0^\prime}  \,
a\left(\nu,\nu^\prime,\mathbf{r_0},\mathbf{r_0^\prime},Y\right)
\phi^{\nu^\prime}\left( \mathbf{r_1},\mathbf{r_2};\mathbf{r_0^\prime} \right).
\label{ansatz}
\end{equation}
Using the translation properties of $\phi^\nu$ and dimensional analysis
we can deduce that the coefficient $a\left(\nu,\nu^\prime,\mathbf{r_0},\mathbf{r_0^\prime},Y\right) $
is of the form
\begin{equation}
a\left(\nu,\nu^\prime,\mathbf{r_0},\mathbf{r_0^\prime},Y\right) \ = \
\tilde{a}(\nu,\nu^\prime,Y)\left[
\left|\mathbf{r_{00^\prime}}\right|^{-(2+2i(\nu-\nu^\prime))}
-(2\pi^2)\delta^2\left(\mathbf{r_{00^\prime}}\right)
\delta\left(\nu-\nu^\prime\right) \right],
\label{eq6}\end{equation}
where the $\delta$-function subtraction has been included to ensure that the diagonal term
from the conformally-invariant part of the kernel is included in the coefficient of the first term,
$a^D(\nu,Y)$.

We therefore rewrite (\ref{ansatz}) as
\begin{eqnarray}
f^\nu(\mathbf{r_1},\mathbf{r_2};\mathbf{r_0}, Y) &
 = & a^D(\nu,Y) \phi^{\nu}\left( \mathbf{r_1},\mathbf{r_2};\mathbf{r_0} \right) \, + \,
 \bar{\alpha}_s \int d\nu^\prime d^2\mathbf{r_0^\prime}
\tilde{a}\left(\nu,\nu^\prime,Y\right)
\nonumber \\
&  & \hspace*{-2cm} \times
\left[
\left|\mathbf{r_{00^\prime}}\right|^{-(2+2i(\nu-\nu^\prime))}
-(2\pi^2)\delta^2\left(\mathbf{r_{00^\prime}}\right) \,
\delta\left(\nu-\nu^\prime\right) \right]
\phi^{\nu^\prime}\left( \mathbf{r_1},\mathbf{r_2};\mathbf{r_0^\prime} \right).
\label{ansatz2}
\end{eqnarray}

Furthermore, we can exploit the translation invariance and scaling covariance of ${\cal K}_1$ to write
the off-diagonal matrix-elements of this kernel as
\begin{eqnarray}
\langle \nu^\prime, \mathbf{r_0}^\prime \left| {\cal K}_1 \right| \nu, \mathbf{r_0} \rangle & \equiv &
 \int \frac{d^2 \mathbf{r_1} d^2 \mathbf{r_2}}{|\mathbf{r_{12}}|^4} d^2 \mathbf{r_3} d^2 \mathbf{r_4}
\phi^{\nu^\prime *}\left(\mathbf{r_1},\mathbf{r_2};\mathbf{r_0}^\prime \right)
{\cal K}_1^{\mathrm{NC}}(\mathbf{r_1}\mathbf{r_2};\mathbf{r_3}\mathbf{r_4})
\phi^{\nu}\left( \mathbf{r_3},\mathbf{r_4};\mathbf{r_0} \right) \nonumber \\ &=& \frac{\chi^{\mathrm{NC}}(\nu,\nu^\prime)}{2\pi^2}
\left|\mathbf{r_{00^\prime}}\right|^{-(2+2i(\nu-\nu^\prime))} ,
\label{eq7}
\end{eqnarray}
where the superscript NC refers to that part of the kernel, ${\cal K}_1$ which is not fully conformal-invariant and therefore
admits off-diagonal matrix-elements.

Inserting the expansion (\ref{ansatz2}) into (\ref{bfkl}) and performing the usual projections, we find
the following expressions for the coefficients $ a^D(\nu,Y)$ and  $\bar{a}(\nu,\nu^\prime,Y)$
up to order $\bar{\alpha}_s$:
\begin{equation}
 a^D(\nu,Y) \ = \ e^{\left(\bar{\alpha_s}\chi_0(\nu)+\bar{\alpha}_s^2\chi_1(\nu)\right) Y}
\label{adiag} \end{equation}

\begin{eqnarray}
\tilde{a}(\nu,\nu^\prime,Y) &= & \frac{1}{2\pi^2}\frac{\bar{\alpha}_s}{\left[1-h(\nu,\nu^\prime)h(\nu,-\nu^\prime)\right]}
\left[\chi^{\mathrm{NC}}(\nu,\nu^\prime)-h(\nu,\nu^\prime) \chi^{\mathrm{NC}}(\nu,-\nu^\prime)\right] \nonumber \\ & \times &
 \frac{\left[e^{\bar{\alpha}_s\chi_0(\nu)Y}-e^{\bar{\alpha}_s\chi_0(\nu^\prime)Y}\right]}
{\left[\chi_0(\nu)-\chi_0(\nu^\prime)\right]}
\label{offdiag} \end{eqnarray}
where
$$ h\left(\nu,\nu^\prime\right) \ = \ 2 \nu \left(\nu-\nu^\prime\right)
\eta^*(\nu^\prime)
\frac{B\left(-i\left(\nu+\nu^\prime\right),2i\nu^\prime\right)}
{B\left(i\left(\nu+\nu^\prime\right),-2i\nu^\prime\right)} $$
with $B(\alpha,\beta)$ being the Euler $\beta$-function,
$$ B(\alpha,\beta) \ \equiv \frac{\Gamma(\alpha)\Gamma(\beta)}{\Gamma(\alpha+\beta)}. $$
The terms involving $h$ arise owing to the fact that the functions
$\phi^\nu$ are not strictly orthonormal and their norm contains a
contributions proportional to $\delta(\nu+\nu^\prime)$.

At first sight it would appear that in order to make progress, it is necessary to determine
the matrix-elements of the non-conformal operator ${\cal K}_1$ between different conformal
eigenfunctions in order to obtain the matrix $\chi(\nu,\nu^\prime)$. This is indeed a daunting prospect!
However, we see that there is a significant simplification if we return to the expansion (\ref{ansatz})
and attempt the integration over the ``centre-of-mass'' $d^2\mathbf{r_0}^\prime$. Before doing this we note that
this integral has an ultraviolet divergence at $\mathbf{r_0}=\mathbf{r_0}^\prime$, which we must regulate
by performing the integral in  $2+2\epsilon$ dimensions and taking the limit $\epsilon \to 0_+$. The sign of
$\epsilon$ will turn out to be crucial to this analysis.

In terms of Feynman parameters $\rho, \, \omega$, we have
\begin{eqnarray}
\int d^{2+2\epsilon}\mathbf{r_0}^\prime  \frac{1}{|\mathbf{r_{00^\prime}}|^{2+2i(\nu-\nu^\prime)}}
 \phi^{\nu^\prime} \left(\mathbf{r_1},\mathbf{r_2};\mathbf{r_0}^\prime\right)
& = &
\frac{\sqrt{2} \nu^\prime}{\pi} \left|\mathbf{r_{12}}\right|^{1+2i\nu^\prime}
\frac{\Gamma\left(1+i(\nu+\nu^\prime\right)}{\Gamma\left(1+i(\nu-\nu^\prime)\right)\Gamma^2\left(1/2+i\nu^\prime\right)}
\nonumber \\ & & \hspace*{-4cm} \times \,
\int_0^1 \frac{d\rho d\omega(1-\rho)\rho^{-1+i(\nu^\prime-\nu)+\epsilon}  [\omega(1-\omega)]^{-1/2+i\nu^\prime}}
{\left[
(1-\rho)\left(\mathbf{r_{10}}^2\omega+\mathbf{r_{20}}^2(1-\omega)\right)+\mathbf{r_{12}}^2\rho \omega(1-\omega)
\right]^{1+i(\nu+\nu^\prime)-\epsilon}}, \label{feynparam}
 \end{eqnarray}
(where we set $\epsilon$ to zero everywhere where it does not affect the ultraviolet singularity).

Examination of the integrand in (\ref{feynparam}) near $\rho=0$
shows that the integration over $\rho$ gives rise to a pole at
$\nu^\prime=\nu+i\epsilon$. This can be used to perform the
integral in (\ref{ansatz2}) over $\nu^\prime$.

At this stage it is more convenient to switch to the mixed representation with momentum transfer $\mathbf{Q}$. Exploiting translation
invariance and omitting an irrelevant phase, we can write the LO eigenfunctions in mixed representation as
\begin{equation}
\tilde{\phi}^{\nu}_{\mathbf{Q}}(\mathbf{r_{12}}) \ = \
\frac{\sqrt{2}\pi}{|\mathbf{r_{12}}|} \int d^2\mathbf{r_0} e^{i \mathbf{Q\cdot r_0}}
\phi^\nu\left(\mathbf{r_1},\mathbf{r_2};\mathbf{r_0}\right),
\end{equation}
which can be written in terms of Bessel functions of complex order (see ref.\cite{np}), as
\begin{equation} 
\tilde{\phi}^{\nu}_{\mathbf{Q}}(\mathbf{r}) \ = \
 2 \eta^*(\nu) \Gamma^2(1-2i\nu) 2^{-6i\nu} \Im m\left\{ J_{i\nu}(Qr)J_{-i\nu}(Qr) \right\},
\end{equation} where $Qr=\mathbf{Q \cdot r}+i|\mathbf{Q \times r}| $.

Taking the Fourier transform of (\ref{ansatz2}) we find

\begin{eqnarray}
\tilde{f}^\nu_{\mathbf{Q}}(\mathbf{r}, Y) &
 = & a^D(\nu,Y) \tilde{\phi}^{\nu}_{\mathbf{Q}} \left( \mathbf{r} \right) \, + \,
 \bar{\alpha}_s \int d\nu^\prime d^2\mathbf{r_{00^\prime}} \,
\tilde{a}\left(\nu,\nu^\prime,Y\right)  e^{i\mathbf{Q\cdot r_{00^\prime}}}
\nonumber \\
&  & \hspace*{-2cm} \times
\left[
\left|\mathbf{r_{00^\prime}}\right|^{-(2+2i(\nu-\nu^\prime))}
-(2\pi^2)\delta^2\left(\mathbf{r_{00^\prime}}\right)
\delta\left(\nu-\nu^\prime\right) \right]
\tilde{\phi}^{\nu^\prime}_{\mathbf{Q}} \left( \mathbf{r} \right).
\label{ansatz3}
\end{eqnarray}

Once again, we can perform the integral over $\mathbf{r_{00^\prime}}$, first promoting
it to $2+2\epsilon$ dimensions in order to regularize the ultraviolet singularity, to get

\begin{eqnarray}
\tilde{f}^\nu_{\mathbf{Q}}(\mathbf{r}, Y) &
 = & a^D(\nu,Y) \tilde{\phi}^{\nu}_{\mathbf{Q}} \left( \mathbf{r} \right) \, + \,
 \bar{\alpha}_s \int d\nu^\prime  \,
\tilde{a}\left(\nu,\nu^\prime,Y\right)  \frac{\Gamma\left(1+i(\nu^\prime-\nu)\right)}{ \Gamma\left(1-i(\nu^\prime-\nu)\right)}
\left(\frac{|\mathbf{Q}|}{2}\right)^{2i(\nu-\nu^\prime)}
\nonumber \\
&  & \hspace*{-2cm} \times
\left[ -i \pi \frac{1}{\left(\nu^\prime-\nu-i\epsilon\right)} \, - \,
\delta\left(\nu-\nu^\prime\right) \right]
\tilde{\phi}^{\nu^\prime}_{\mathbf{Q}} \left( \mathbf{r} \right),
\label{solution}
\end{eqnarray}
and note the pole at $\nu^\prime=\nu+i\epsilon$ \footnote{There are other poles in the upper plane where $\nu^\prime=\nu+in$
for integer $n$. These correspond to contributions from conformal wavefunctions with higher conformal spin, which we are
neglecting since the rapidity dependence is substantially reduced for these eigenfunctions. They can be consistently eliminated
by integrating over the azimuthal angle between $\mathbf{Q}$ and $\mathbf{r}$.}

We now split the function $\tilde{\phi}^{\nu^\prime}_{\mathbf{Q}} \left( \mathbf{r} \right)$ into two parts
\begin{equation}
\tilde{\phi}^{\nu^\prime}_{\mathbf{Q}} \left( \mathbf{r} \right) \ = \ \tilde{\phi}^{\nu^\prime}_{\mathbf{Q}} \left( \mathbf{r} \right)^{[1]} \, + \,
\tilde{\phi}^{\nu^\prime}_{\mathbf{Q}} \left( \mathbf{r} \right)^{[2]}, \end{equation}

where
\begin{equation}
\tilde{\phi}^{\nu^\prime}_{\mathbf{Q}} \left( \mathbf{r} \right)^{[1]} \ = \ \frac{\sqrt{2}\nu^\prime}{\pi^2}
\int_{{\cal R}_1} d^2\mathbf{R} \left[ \frac{|\mathbf{r}|}{|\mathbf{R}+\mathbf{r}/2||\mathbf{R}-\mathbf{r}/2|}\right]^{1+2i\nu^\prime}
\end{equation}
where the region indicated by ${\cal R}_1$ is the region of $\mathbf{R}$ for which
$$ |\mathbf{Q}| \ >  \frac{2 |\mathbf{r}|}{|\mathbf{R}+\mathbf{r}/2||\mathbf{R}-\mathbf{r}/2|} $$
and $\tilde{\phi}^{\nu^\prime}_{\mathbf{Q}} \left( \mathbf{r} \right)^{[2]}$ is the integral over the remaining region, ${\cal R}_2$.

In region ${\cal R}_1$, the integral over $\nu^\prime$ is performed by closing the contour in the upper-half plane, thereby
picking up the pole, whereas in region    ${\cal R}_2$, the contour is closed in the lower half-plane, thereby missing
the pole. Accounting for the $\delta$-function subtraction we arrive at the final result
\begin{eqnarray}
\tilde{f}^\nu_{\mathbf{Q}}(\mathbf{r}, Y)  &
 = & a^D(\nu,Y) \tilde{\phi}^{\nu}_{\mathbf{Q}} \left( \mathbf{r} \right) \, - \,
  (2\pi)^2 \, \bar{\alpha}_s  \,
\tilde{a}\left(\nu,\nu, Y\right) \tilde{\phi}^{\nu}_{\mathbf{Q}} \left( \mathbf{r} \right)^{[2]}
\nonumber \\ & = &
 e^{(\bar{\alpha_s}\chi_0(\nu)+\bar{\alpha_s}^2\chi_1(\nu))Y} \tilde{\phi}^{\nu}_{\mathbf{Q}} \left( \mathbf{r} \right) \, - \,
 \bar{\alpha}_s^2  \, Y \chi^{\mathrm{NC}}(\nu,\nu)
 \tilde{\phi}^{\nu}_{\mathbf{Q}} \left( \mathbf{r} \right)^{[2]}
 \label{soln2}
 \end{eqnarray}

The problem is therefore reduced to the determination of the diagonal matrix-elements of that part of the NLO Kernel
which is {\it not} covariant under the full set of M\"{o}bius transformations. We can extract this part of the kernel
by writing
\begin{equation} {\cal K}_1^{\mathrm{NC}}(\{\mathbf{r_i}\}) \ = \
 \frac{1}{2} \left\{  {\cal K}_1(\{\mathbf{r_i}\})-\frac{1}{|\mathbf{r_3}|^4|\mathbf{r_4}|^4} {\cal K}_1(\{1/\mathbf{r_i}\}) \right\}.
\label{sub}
\end{equation}
We see immediately that this expression vanishes for any part of the kernel which is covariant under inversions.
We can furthermore project the diagonal term $\chi^{\mathrm{NC}}(\nu,\nu)$, from eq.(\ref{eq7}) by integrating
both sides over $\mathbf{r_0}$, which leads to
\begin{eqnarray}
\chi^{\mathrm{NC}}(\nu,\nu) & = & -i\pi^2 \frac{\eta^*(\nu)}{\nu}
 \lim_{\nu\to\nu^\prime} \left(\nu^\prime-\nu\right)
\int \frac{d^2 \mathbf{r_1}d^2 \mathbf{r_2}d^2 \mathbf{r_3}d^2 \mathbf{r_4}}{|\mathbf{r_{12}}|^4}
 \left(|\mathbf{r_1}|\mathbf{r_2}|\right)^{1+2i\nu^\prime}
\nonumber \\ & &   \hspace*{2cm}
  \phi^{\nu^\prime}(\mathbf{r_1},\mathbf{r_2};\mathbf{0})
 {\cal K}_1^{\mathrm{NC}}(\mathbf{r_1}, \mathbf{r_2}, \mathbf{r_3}, \mathbf{r_4})
 \phi^{\nu}(\mathbf{r_3},\mathbf{r_4};\mathbf{0})
\end{eqnarray}
After applying a suitable change of variables for the subtraction term in (\ref{sub}), this may be rewritten as
 \begin{eqnarray}
\chi^{\mathrm{NC}}(\nu,\nu) & = & -i\frac{\pi^2}{2} \frac{\eta^*(\nu)}{\nu}
 \lim_{\nu\to\nu^\prime} \left(\nu^\prime-\nu\right)
\int \frac{d^2 \mathbf{r_1}d^2 \mathbf{r_2}d^2 \mathbf{r_3}d^2 \mathbf{r_4}}{|\mathbf{r_{12}}|^4}
 \left[\left(|\mathbf{r_1}|\mathbf{r_2}|\right)^{1+2i\nu^\prime}
 -\left(|\mathbf{r_3}|\mathbf{r_4}|\right)^{1+2i\nu}
\right]
\nonumber \\ & &   \hspace*{2cm}
  \phi^{\nu^\prime}(\mathbf{r_1},\mathbf{r_2};\mathbf{0})
 {\cal K}_1(\mathbf{r_1}, \mathbf{r_2}, \mathbf{r_3}, \mathbf{r_4})
 \phi^{\nu}(\mathbf{r_3},\mathbf{r_4};\mathbf{0})
\end{eqnarray}

As an example, we take
\begin{equation}
{\cal K}_1(\mathbf{r_1}, \mathbf{r_2}, \mathbf{r_3}, \mathbf{r_4}) \ = \
 \frac{1}{(4\pi^2)} \frac{\mathbf{r_{24}} \cdot \mathbf{r_{12}}}
{\left(|\mathbf{r_{13}}| |\mathbf{r_{24}}| |\mathbf{r_{34}}|\right)^2} \ln\left(\frac{\mathbf{r_{14}}^2}{\mathbf{r_{34}}^2}\right),
\end{equation}
which is one of the terms of ${\cal K}_1$ given in \cite{ffp}. The integrations over $\mathbf{r_i}$ may be performed
by employing a judicious ordering for the integrations, yielding
\begin{equation}
\chi^{\mathrm{NC}}(\nu,\nu) \ = \ \frac{2\nu}{(1+4\nu^2)} \Im m \{ \Psi^\prime\left(1/2+i\nu\right) \}
\end{equation}
A great deal of work is still required in order to determine this quantity for all the other terms
of ${\cal K}_1$ in \cite{ffp}. Nevertheless we have now shown that the problem is, at least
in principle, tractable.

Analytic approximations for $\tilde{\phi}^{\nu}_{\mathbf{Q}} \left( \mathbf{r} \right)^{[2]}$ can be obtained for
the limits of very large or very small momentum transfer, $\mathbf{Q}$.

For $\mathbf{Q\cdot r} \, \ll \, 1$, in the region ${\cal R}_2$, we have $|\mathbf{R}| \, \gg \, |\mathbf{r}| $
so that
\begin{equation}
  \tilde{\phi}^{\nu}_{\mathbf{Q}} \left( \mathbf{r} \right)^{[2]} \ \approx \ \frac{2\sqrt{2}\nu}{\pi}
 16^{i\nu} |\mathbf{r}|^{-2i\nu} (\mathbf{Q\cdot r})^{4i\nu} \int_1^\infty \frac{dz}{z^{2+2i\nu}} J_0\left(\sqrt{2\mathbf{Q.r}z}\right),
\end{equation}
and if  $\mathbf{Q\cdot r}$ is sufficiently small this becomes
\begin{equation}
 \tilde{\phi}^{\nu}_{\mathbf{Q}} \left( \mathbf{r} \right)^{[2]} \ \approx \ \frac{2\sqrt{2}\nu}{\pi}
 16^{i\nu} |\mathbf{r}|^{-2i\nu} (\mathbf{Q\cdot r})^{4i\nu}\end{equation}
This does not vanish as $|\mathbf{Q}| \, \to \, 0$, but oscillates very rapidly as $\nu$ is varied over a small range,
 so that its effect on any realistic  amplitude, which involves an integral over $\nu$ will vanish. However, {\it caveat lector}, for small $\nu$,
 there is a very small range, $\Delta \nu$, of $\nu$ over which the eigenvalues, $\chi_0, \, \chi_1$ do not vary substantially, which means
that in order for these oscillations to lead to a negligible correction we need to go to sufficiently small $\mathbf{Q}$ such that
 $$ |\mathbf{Q}| \ \ll \ |\mathbf{r}| \, e^{-1/\delta\nu}. $$

On the other hand, if  $\mathbf{Q\cdot r} \, \gg \, 1$, then the region ${\cal R}_2$ dominates region  ${\cal R}_1$
in the Fourier transform of $\phi^\nu$, so that in this case we have
\begin{equation}
\tilde{\phi}^{\nu}_{\mathbf{Q}} \left( \mathbf{r} \right)^{[2]}  \ \approx \
\tilde{\phi}^{\nu}_{\mathbf{Q}} \left( \mathbf{r} \right). \end{equation}

\begin{figure}[h]
\epsfig{file=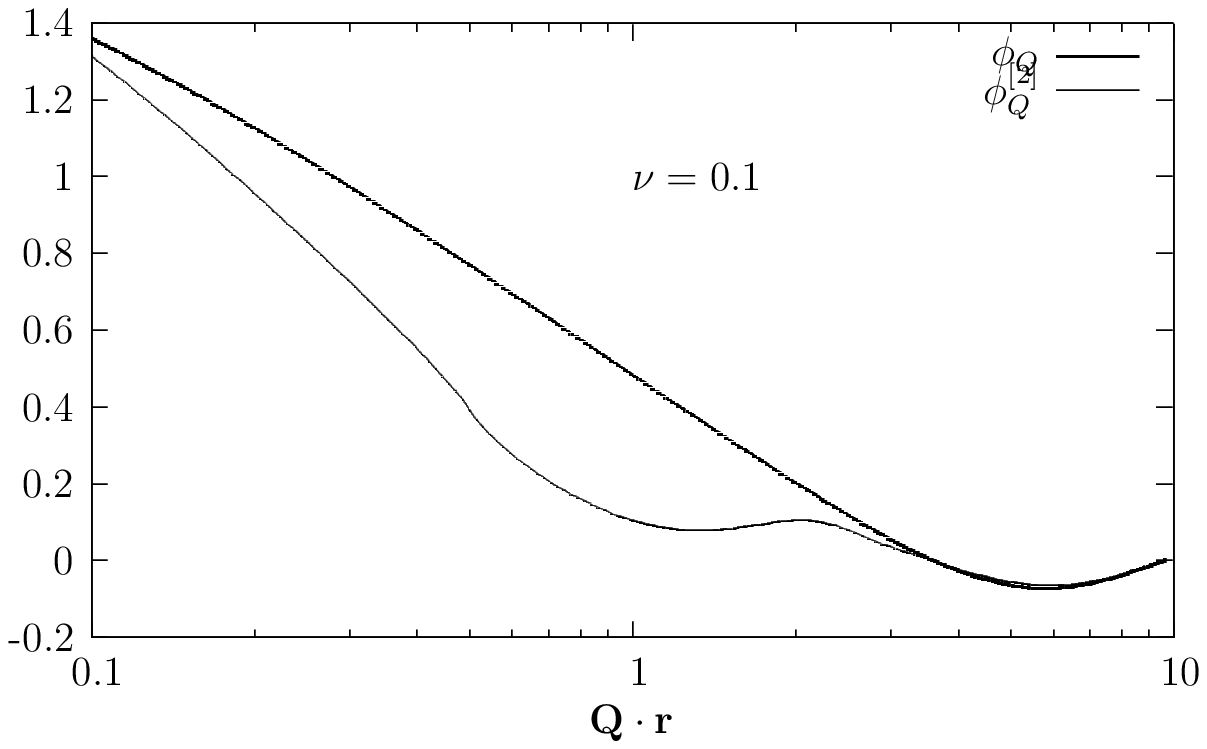,width=5.3cm}
\epsfig{file=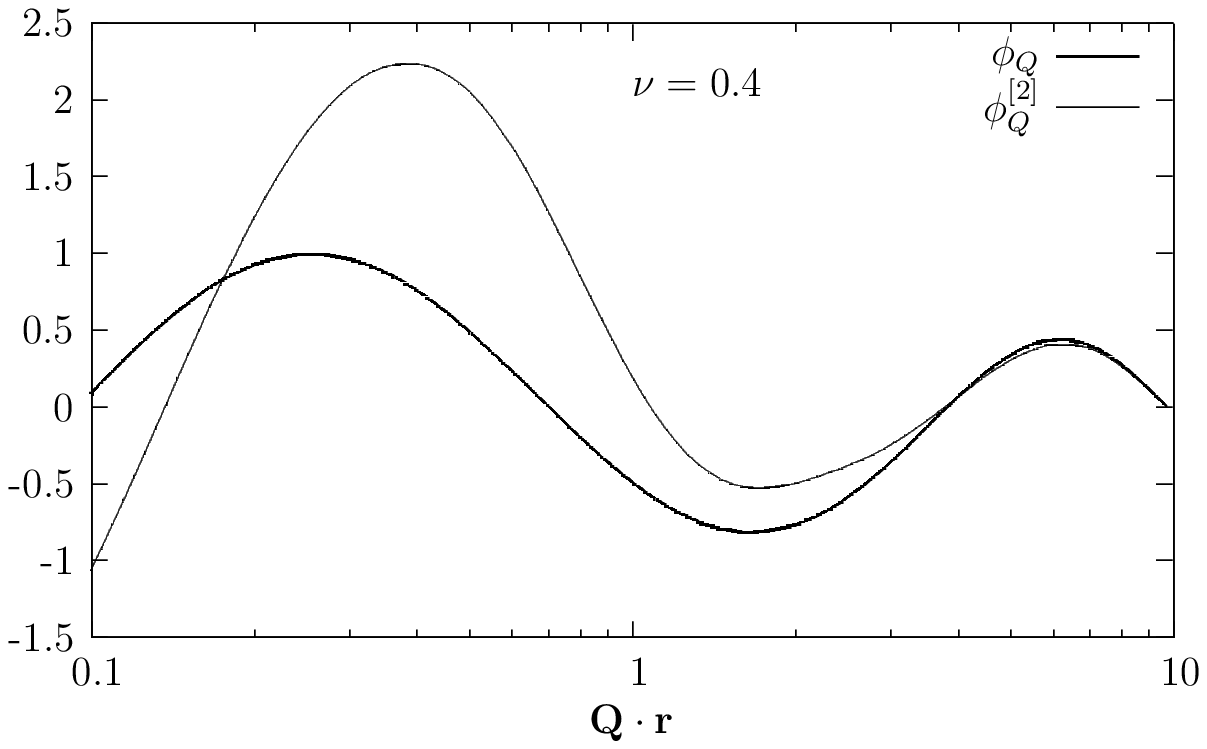,width=5.3cm}
\epsfig{file=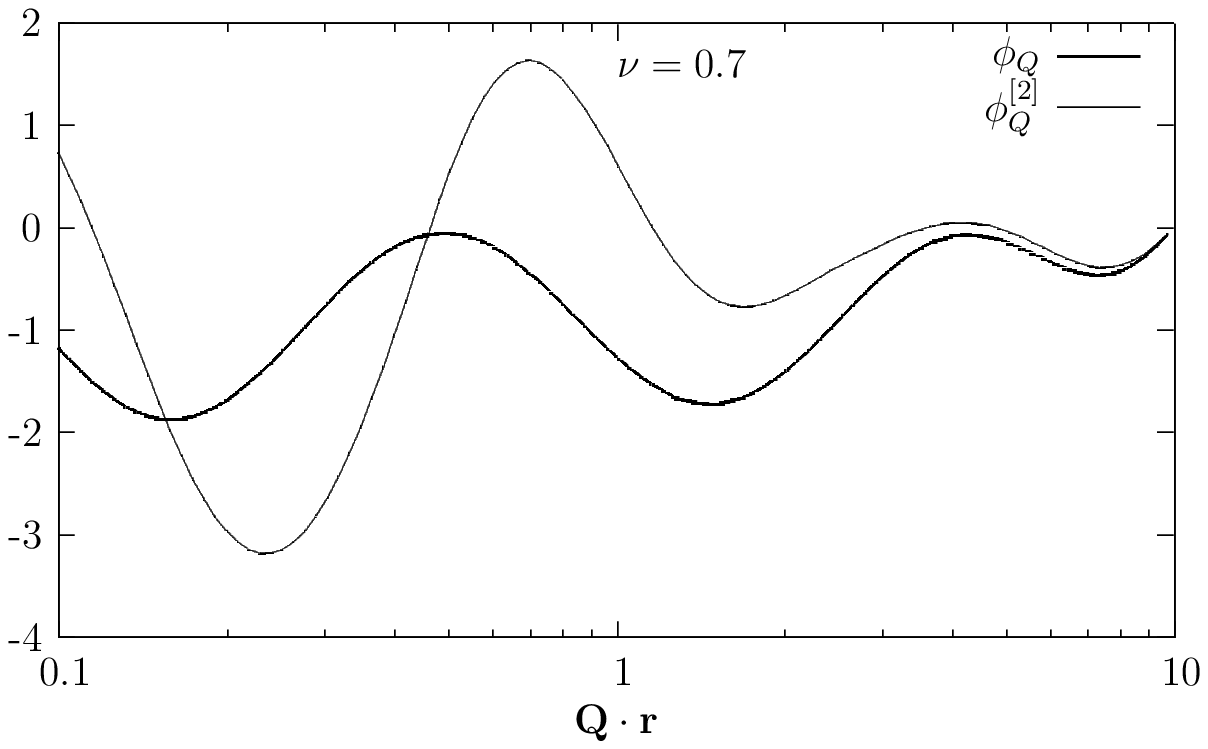,width=5.3cm}
\caption[]{Plots of the wavefunction
 $\tilde{\phi}^{\nu}_{\mathbf{Q}} \left( \mathbf{r} \right)^{[2]}$
from region ${\cal R}_2$ and the total wavefunction, 
 $\tilde{\phi}^{\nu}_{\mathbf{Q}} \left( \mathbf{r} \right)$,
against  $\mathbf{Q \cdot r}$ for different values of $\nu$. In these
and subsequent plots, a factor of $|\mathbf{r}|^{-2i\nu}$ 
has been factored out.}
\label{fig1}
\end{figure}

\begin{figure}[h]
\epsfig{file=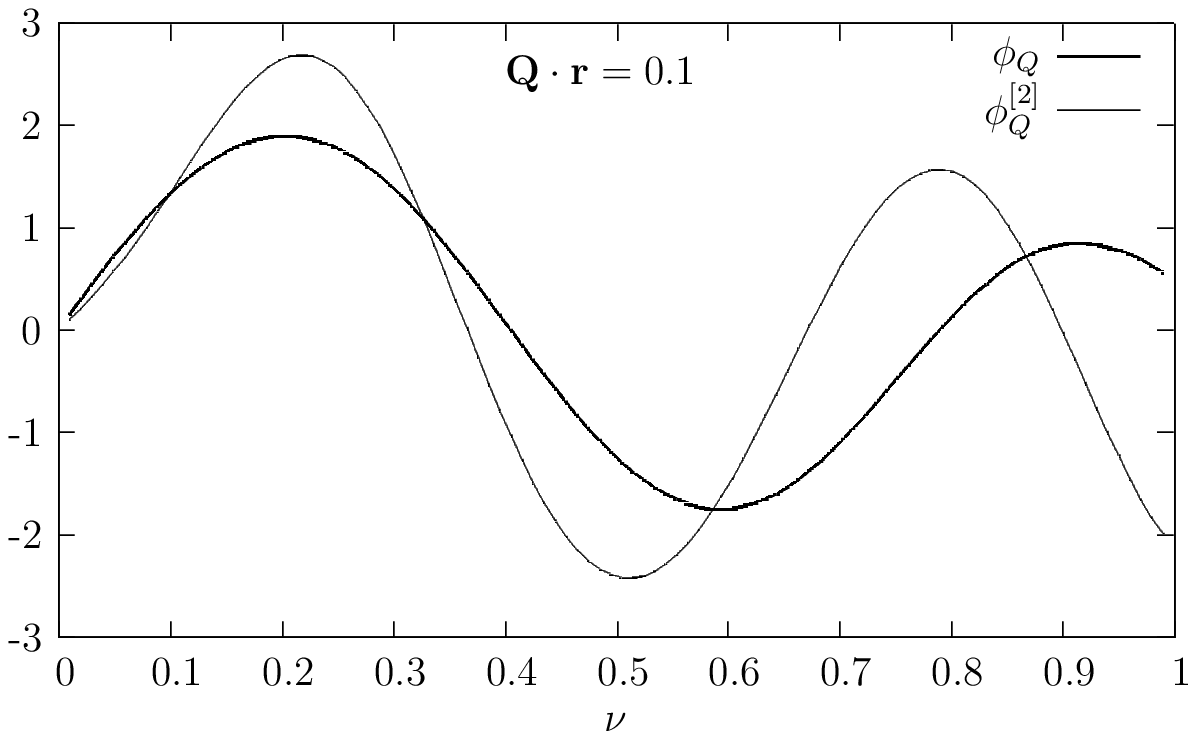,width=5.3cm}
\epsfig{file=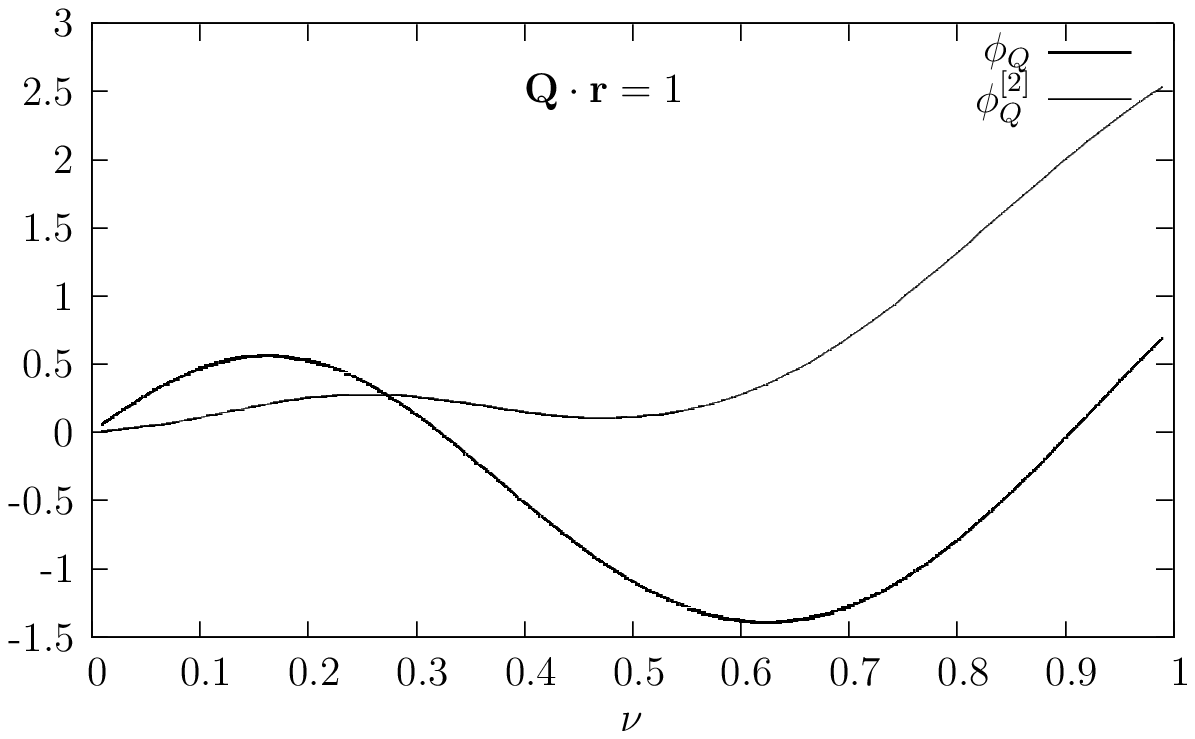,width=5.3cm}
\epsfig{file=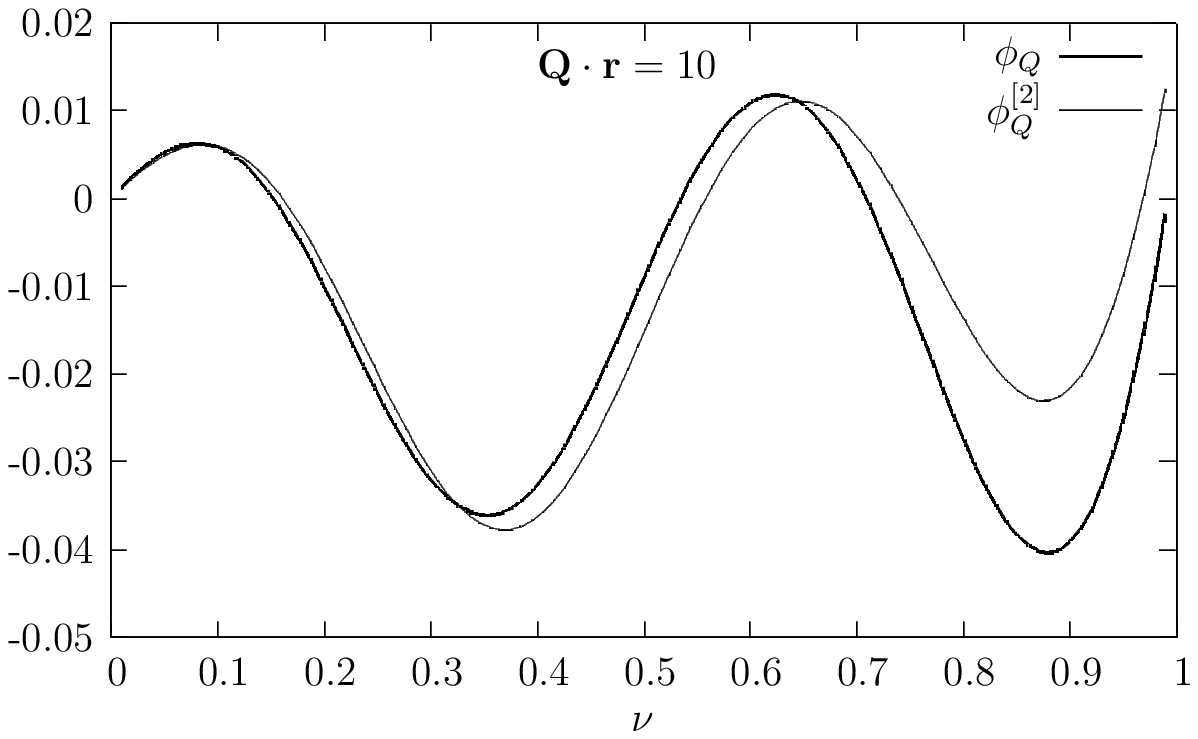,width=5.3cm}
\caption[]{Plots of the wavefunction
 $\tilde{\phi}^{\nu}_{\mathbf{Q}} \left( \mathbf{r} \right)^{[2]}$
from region ${\cal R}_2$ and the total wavefunction, 
 $\tilde{\phi}^{\nu}_{\mathbf{Q}} \left( \mathbf{r} \right)$,
against $\nu$ for low, medium and high, values of $\mathbf{Q \cdot r}$}
\label{fig2}
\end{figure}

For intermediate values of momentum transfer, where $\mathbf{Q
\cdot r} \, \sim \, 1$, we need to resort to numerical
approximations for the integral in region ${\cal R}_2$. In
Fig.\ref{fig1}, we show the real parts of the wavefunctions
$\tilde{\phi}^{\nu}_{\mathbf{Q}} \left( \mathbf{r} \right)^{[2]}$
and $\tilde{\phi}^{\nu}_{\mathbf{Q}} \left( \mathbf{r} \right)$
against $\mathbf{Q\cdot r}$ in this region for three different
values of $\nu$, and in Fig.\ref{fig2} we show the same quantities
against $\nu$ in the region $0 \, < \, \nu \, < 1$ for three
different values of $\mathbf{Q\cdot r}$. We see from these that as
explained above the two functions almost coincide for sufficiently large
$|\mathbf{Q}|$ and that for small values of $|\mathbf{Q}|$, there
are more rapid oscillations with $\nu$.

In summary, we have developed an algorithm for calculating the
non-forward BFKL wavefunction up to corrections of order
$\bar{\alpha}_s$ allowing for an NLO kernel which is not covariant
under the full M\"{o}bius transformations.

\section*{Acknowledgment}

The grateful to Agustin Sabio-Vera for useful conversations, and the Theory Division at CERN for
its hospitality.

\end{document}